\documentclass[letterpaper]{article} 
\usepackage[]{aaai24}  
\usepackage{times}  
\usepackage{helvet}  
\usepackage{courier}  
\usepackage[hyphens]{url}  
\usepackage{graphicx} 
\usepackage{natbib}  
\usepackage{caption} 
\usepackage{amsfonts}
\usepackage{amssymb} 
\usepackage{subfig}
\usepackage{algorithm}
\usepackage{algpseudocode}
\usepackage{xcolor}
\usepackage{amsmath}
\usepackage{multirow}
\usepackage{booktabs}
\usepackage{xfrac}
\usepackage{wasysym}
\usepackage{graphicx}
\usepackage{array}
\usepackage{dashrule}
\usepackage{tikz}
\usepackage{enumitem}
\usepackage{microtype}

\usetikzlibrary{decorations.markings}

\usepackage{xcolor}
\newcommand{\answerYes}[1]{\textcolor{blue}{#1}} 
\newcommand{\answerNo}[1]{\textcolor{teal}{#1}} 
\newcommand{\answerNA}[1]{\textcolor{gray}{#1}}

\newcommand{\PreserveBackslash}[1]{\let\temp=\\#1\let\\=\temp}
\newcolumntype{L}[1]{>{\PreserveBackslash\raggedright}p{#1}}
\newcolumntype{C}[1]{>{\PreserveBackslash\centering}p{#1}}

\usepackage{newfloat}
\usepackage{listings}
\DeclareCaptionStyle{ruled}{labelfont=normalfont,labelsep=colon,strut=off} 
\floatstyle{ruled}
\newfloat{listing}{tb}{lst}{}
\floatname{listing}{Listing}
\title{Using Causality to Infer Coordinated Attacks in Social Media}
\author{
    Isura Manchanayaka,
    Zainab Razia Zaidi,
    Shanika Karunasekera,
    Christopher Leckie
}
\affiliations{
    The University of Melbourne, Australia\\

    imanchanayak@student.unimelb.edu.au,
    \{zainab.raziazaidi,
    karus,
    caleckie\}@unimelb.edu.au
}

\usepackage{bibentry}

\begin{document}

\maketitle

\begin{abstract}
    The rise of social media has been accompanied by a dark side with the ease of creating fake accounts and disseminating misinformation through coordinated attacks. Existing methods to identify such attacks often rely on thematic similarities or network-based approaches, overlooking the intricate causal relationships that underlie coordinated actions. This work introduces a novel approach for detecting coordinated attacks using Convergent Cross Mapping (CCM), a technique that infers causality from temporal relationships between user activity. We build on the theoretical framework of CCM by incorporating topic modelling as a basis for further optimizing its performance. We apply CCM to real-world data from the infamous IRA attack on US elections, achieving F1 scores up to 75.3\% in identifying coordinated accounts. Furthermore, we analyse the output of our model to identify the most influential users in a community. We apply our model to a case study involving COVID-19 anti-vax related discussions on Twitter. Our results demonstrate the effectiveness of our model in uncovering causal structures of coordinated behaviour, offering a promising avenue for mitigating the threat of malicious campaigns on social media platforms.

\end{abstract}

\section{Introduction}
\label{sec:introduction}
While social media platforms have witnessed explosive growth due to factors like peer pressure, evolving communities, and influencer culture, this increased engagement has fueled a parallel threat: the ease of generating fake accounts has increased the spread of misinformation and disinformation. Politically driven campaigns, seeking to manipulate public opinion and achieve specific goals, rely on large numbers of coordinated accounts to amplify their messages and maximize the impact. Existing techniques primarily focus on identifying coordinated behaviours based on thematic similarities, overlooking the intricate causality relationships that underlie coordinated actions. This complexity in detecting true coordination represents a significant gap in current methodologies, forming the core motivation for our work.

There have been numerous examples of coordinated attacks on social media. One of the most notable instances was the influence of Russia's IRA on the USA presidential elections via Twitter and Facebook \citep{roberts.muellerUNITEDSTATESAMERICA2018}. The \citeauthor{permanentselectcommitteeonintelligenceExposingRussiaEffort2018} identified 3,841 coordinated Twitter accounts and 470 Facebook pages that were affiliated with the IRA in 2017. In 2018, Twitter publicly released tweets and users related to this case. In 2019, the UK general elections were influenced by coordinated users who polarized political opinions on Twitter \citep{nizzoliCoordinatedBehaviorSocial2021}.

There have been numerous attempts to identify coordination in Online Social Networks (OSNs). Some work in this field sought to identify campaigns in social media \citep{leeContentdrivenDetectionCampaigns2011, leeDetectingCollectiveAttention2012}. A major limitation of their work is the assumption that coordination is reflected in the ``theme'' of messages while other aspects of behaviour are ignored. Network-based approaches \citep{pachecoUncoveringCoordinatedNetworks2021a, nizzoliCoordinatedBehaviorSocial2021, weberAmplifyingInfluenceCoordinated2021, magelinskiSynchronizedActionFramework2021, hristakievaSpreadPropagandaCoordinated2022} tend to define coordination in terms of community detection on user similarity graphs. \citet{weberAmplifyingInfluenceCoordinated2021} highlighted several coordination strategies: \emph{pollution} – flooding a community with repeated content, \emph{boost} – heavily reposting content to make the topic appear popular, and \emph{bully} – groups collectively harassing another individual or a community. In contrast, \citet{zhangVigDetKnowledgeInformed2021} and \citet{sharmaIdentifyingCoordinatedAccounts2021} define coordination in terms of the synchronicity of users over time. They try to identify coordinated users using masked self-attention \citep{vaswani2017attention} to encode the event history, using an approach similar to the prediction model for marked temporal point processes by \citet{shchurIntensityFreeLearningTemporal2020}. Network-based and theme-based approaches work under the assumption that the content is the governing factor of the coordinated behaviour. The activity-based approaches work under the assumption that active times of coordinated users are causally linked with each other. In contrast, we consider that coordination should be reflected in the user activity traces.

We propose to identify how influence flows within a community of users by assessing causality between pairs of users by exploring the layered dynamics and dependencies between users. Causality offers a nuanced understanding of the users who influence or trigger coordinated responses from others. The idea of causality not only enhances the precision of coordination detection, but also provides a deeper understanding of the mechanisms driving coordinated activities.

Inferring causal structures through activity traces is more reliable than focusing on textual data for several reasons. First, activity traces provide a record of user interactions, reducing the risk of misinterpretation in text analysis. Second, textual content is often context-dependent and subject to nuances that can obscure true causal relationships, making it challenging to accurately discern influence and coordination. In contrast, activity traces reflect the timing and sequence of actions, offering a clearer picture of how influence propagates through a network, allowing for more precise and robust detection of coordinated behaviours.

In order to address the problem of inferring causality between users as a basis for identifying coordination, we build on the theory of Convergent Cross Mapping (CCM) \cite{sugiharaDetectingCausalityComplex2012}. CCM is a powerful technique that has been used to identify causality in applicaitons such as ecology and climatology. However, utility of CCM for inferring coordinated behaviour among social media users has not been considered in the literature.


In this article, we investigate methods of identifying coordination using convergent cross mapping, and evaluate the performance of our model on real data. Our research aims to address the following questions:

\begin{enumerate}[label=RQ\arabic*., align=left]
  \item How effective is inferring coordination using causal structures of users?
  \item Based on semantics, what methods can be implemented to optimize the performance of our model?
  \item What are the key limitations and challenges associated with inferring coordination using causal structures?
\end{enumerate}

Our experiments on the IRA dataset \cite{permanentselectcommitteeonintelligenceExposingRussiaEffort2018, roberts.muellerUNITEDSTATESAMERICA2018} show that cross mapping each pair of users can identify coordinated pairs of users accurately. Moreover, the coordinated users who were identified by our model belong to clearly separated clusters of interests. We achieve F1 scores up to 75.3\%. Further, we exploit the clustered nature of users to optimize our model.

\newcommand*{\xdash}[1][3em]{\rule[0.5ex]{#1}{0.7pt}}
\definecolor{crimson}{HTML}{dc143c}
\definecolor{dodgerblue}{HTML}{2090ff}

\newcommand{\xddash}{\hbox to 2em{\leaders\hbox to 3pt{\hss . \hss}\hfil}}

\begin{figure*}[t!]
  \centering
  \subfloat[\centering \label{fig:sim_1}]{{\includegraphics[width=0.32\linewidth]{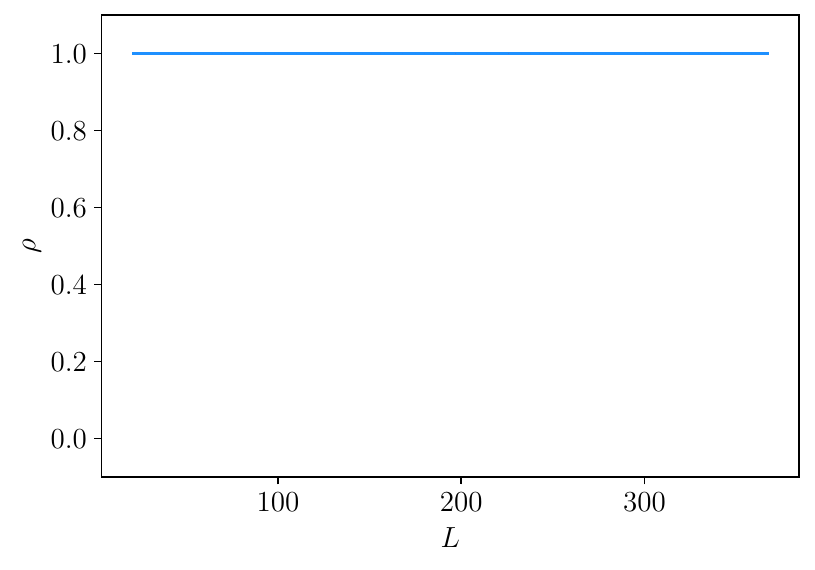} }}%
  \subfloat[\centering \label{fig:sim_2}]{{\includegraphics[width=0.32\linewidth]{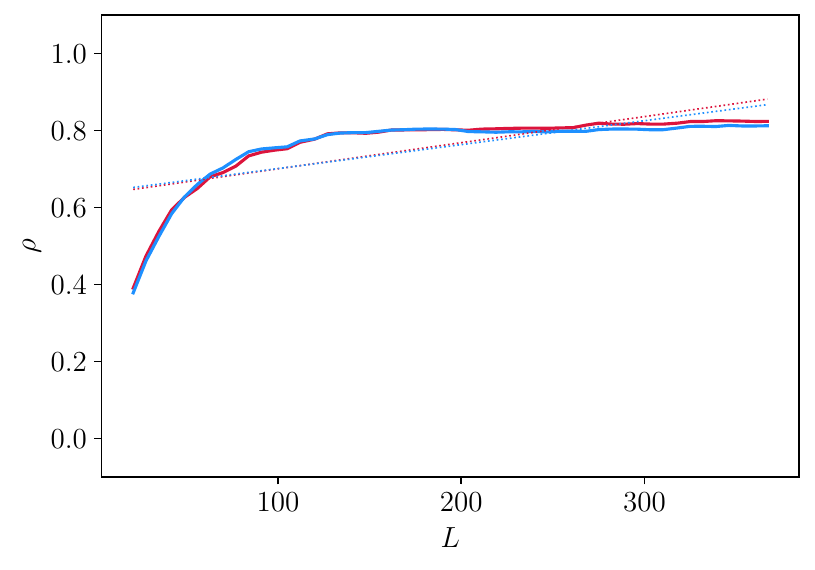} }}%
  \par\medskip
  \subfloat[\centering \label{fig:sim_3}]{{\includegraphics[width=0.32\linewidth]{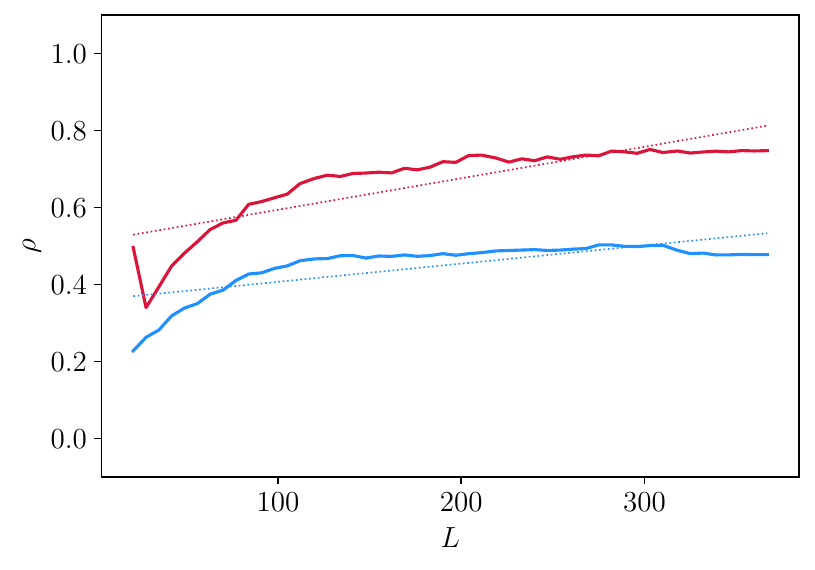} }}%
  \subfloat[\centering \label{fig:sim_4}]{{\includegraphics[width=0.32\linewidth]{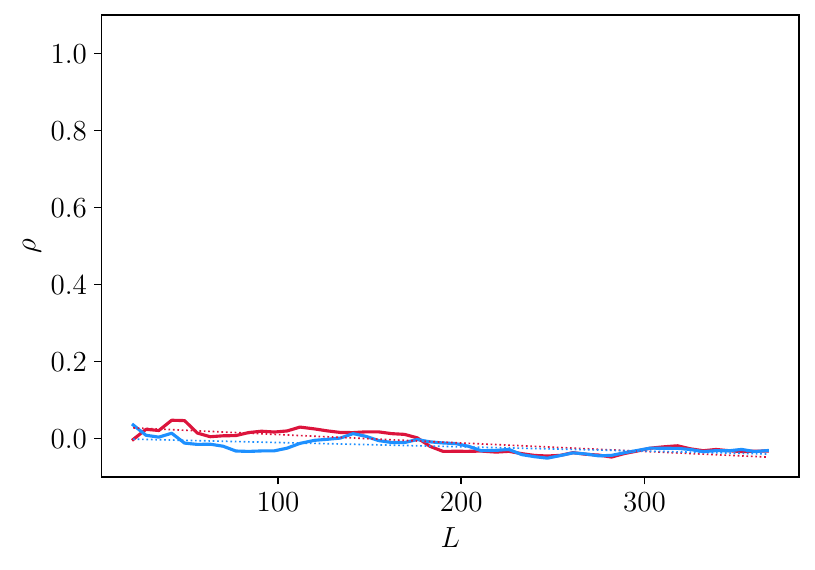} }}%
  \caption{Motivating example of the use of CCM to model causal behaviour in simulated social media data. We show the variation of the correlation of predictions (vertical axis denoted $\rho$) for prediction about two simulated users $u_1$ and $u_2$, where $u_2$ follows $u_1$ on social media, as the library lengths $L$ (i.e., sample periods) increase. CCM implies causation if the correlation is increasing for increasing library lengths. \textcolor{crimson}{\xdash[2em]} :  predictions for $u_1$ given $u_2$'s shadow manifold i.e., history, \textcolor{dodgerblue}{\xdash[2em]} : predictions for $u_2$ given $u_1$'s shadow manifold,  \raisebox{0.4ex}{\textcolor{crimson}{\xddash}}: linear regression drawn for $u_1$'s variation of correlation, and \raisebox{0.4ex}{\textcolor{dodgerblue}{\xddash}}: linear regression drawn for $u_2$'s variation of correlation. (a) $u_2$ posts after $u_1$, who posts at regular intervals. (b) $u_2$ posts after $u_1$, who posts at irregular intervals. (c) $u_2$ posts after $u_1$, who posts at irregular intervals. However, $u_2$ also posts at random times without $u_1$ triggering $u_2$'s behaviour. (d) $u_1$ and $u_2$ behaves randomly.}%
  \label{fig:simulations}%
\end{figure*}






\section{Background and Problem Statement}
\label{sec:problem-statement}

\subsubsection{Convergent Cross Mapping.}

Unraveling relationships within complex systems often leads researchers to study nuanced separation between correlation and causation. While correlation signifies a statistical association between two variables, it falls short of establishing a cause-and-effect relationship. In contrast, causation implies a direct influence of one variable on another, suggesting a deeper understanding of the underlying mechanisms governing a system. Convergent Cross Mapping (CCM) \cite{sugiharaDetectingCausalityComplex2012} is a powerful technique that can determine causality using the variation of correlation at different training sample sizes (known  as library lengths) of predictions. CCM uses Takens' principle \cite{Takens} to detect if two variables belong to the same dynamic system. Consider two time series variables $X$ and $Y$. CCM establishes the causality between variables by examining the predictive accuracy of a cross-mapped model built using historical Y data to reconstruct X states. Causality is suggested by the convergence of these reconstructed states towards the actual X values.

CCM has been primarily found application in ecology and \cite{sugiharaDetectingCausalityComplex2012, Clark2015SpatialCC, Ye2015DistinguishingTC, FROSSARD2018136} climatology \cite{Nes2015CausalFI, Luo2015QuestionableDE, Zhang2018DynamicalEF}. CCM was reviewed and provided improvements in works of \citet{Ye2015DistinguishingTC, Krakovsk2016TestingFC} and \citet{Tsonis2018}. The study in \citet{Cobey2016LimitsTC} explores the limitations of CCM such as its sensitivity to periodicity. We consider that there is no reason for there to be such periodical fluctuations of activity of users in OSNs, but only major events govern the activity of users. CCM has not been widely studied in the context of social networks other than the work of \citet{causal6975587} and \citet{CHUNG2020103108}. However, their work is not geared to infer coordination using causal relationships, but instead to confirm their results with alternative network measures. Thereby, we identify the gap in literature that CCM has yet to be applied in social media contexts to determine coordinating behaviours. 


\subsubsection{Preliminaries.}

We refer to an interaction made by a user in the OSN as an \emph{event}. For example, on Twitter (now called X), a tweet authored by a user is considered an event. The set of timestamps of events authored by a user $u$ is called an \emph{activity trace} $\{t_{0,u}, t_{1,u}, \dots\}$. \emph{Influence flow} is defined as the causal effect one user's activity has on another user's activity. If there is an influence flow from user $u_1$ to $u_2$, we say $u_1 \Rightarrow u_2$. On a higher level, if $u_1 \Rightarrow u_2$, then there is a high tendency that $u_2$ gets active after $u_1$'s action. If $u_1 \Rightarrow u_2$ and $u_2 \Rightarrow u_1$, we say there is a \emph{bidirectional coupling}. If $u_1 \Rightarrow u_2$ but $u_1 \nRightarrow u_2$, we say that there exists a \emph{unidirectional coupling}. We named the directed graph where the vertices are users and the edges are influence flows to be the \emph{influence graph} for the sample of users we consider.

Let $U$ be a set of users in an OSN. Say we determine a time period $(t_{\text{start}},t_{\text{end}})$ that presumably contains coordinated anomalous activities based on observations. Let $T$ be the set of activity traces performed by each user in $U$ in the above time interval.

\noindent\textbf{%
Problem Definition.
}\emph{Given a dataset of activity traces $T$, find pairs of users that are causally influenced unidirectionally or bidirectionally by measuring their prediction scores through Convergent Cross Mapping. Find the users that belong to such influencing pairs and mark them as coordinating users.}

\section{Methodology}
\label{sec:methodology}
\subsubsection{Motivating Example.}

Our results for simulated users highlight the applicability of CCM for social media data. We simulated two users $u_1$ and $u_2$ to model different stages of a simple leader-follower behavioural spectrum. We assumed that the extremes of this spectrum to be: (1) the follower can only be activated once with a lag after the leader is active, (2) the behaviour is random for both agents.  We applied CCM to observe cross map prediction accuracy measured with correlation at different library lengths. Figures \ref{fig:sim_1}, \ref{fig:sim_2}, \ref{fig:sim_3}, and \ref{fig:sim_4} show our simulation results. The increasing nature of correlation when there is a leader-follower behaviour motivated us to adapt CCM to analyse on real online social network (OSN) data.

\subsubsection{Model.}

Say we analyse a set of users' ($U$) activity in a time period $(t_{\text{start}}, t_{\text{end}})$. First, we record timestamps of events authored by each user $u\in U$ as $T_u = \{t_{0,u}, t_{1,u}, t_{2,u}, \dots\}$. Subsequently, every $T_u$ is vectorized to a fixed size $L$, $X_u=\langle X_u(1), X_u(2), X_u(3), \dots, X_u(L) \rangle$. Essentially, we partition the time series into bins of size $I=\sfrac{(t_{\text{end}} - t_{\text{start}})}{L}$. Here, $X_u(a)=|\{ t \mid t \in T_u, (a - 1)I \leq t < aI \}|$.  The embedding function $e: U \times \mathbb{Z}_{\geq 0} \rightarrow \mathbb{Z}_{\geq 0}^E$ transforms each time series into a series of lagged-coordinate embeddings. For a lag $\tau > 0$ and embedding size $E > 1$, a point in $X_u$ at time $t$ is transformed as $e(u, t)=\langle X_u(t), X_u(t - \tau), \dots, X_u(t - (E - 1)\tau)\rangle$. This embedding results in a manifold $M_u=\left [e(u, 1), e(u, 2), \dots, e(u, L) \right ]$ for each user $u$. For a unique pair of users $u_1$ and $u_2$, $M_{u_1}$ and $M_{u_2}$ can be considered as two shadow manifolds for the attractor manifold of the original behaviour system of these two users given by $M_{u_1, u_2}=[\langle X_{u_1}(t), X_{u_2}(t) \rangle \mid t = 1, 2, \dots, L]$. We now cross map $X_{u_1}$ using $M_{u_2}$ and vice versa. Specifically, we use a nearest neighbors model with $k=E + 1$. Unseen data in a future time window is then tested with the fitted model to obtain predictions $\hat{X}_{u_1}\mid M_{u_2}$ and $\hat{X}_{u_2}\mid M_{u_1}$. The cross-correlation $\rho$ of each prediction $\hat{X}_{u_1}\mid M_{u_2}$ is compared with ground truth $X_{u_1}$ for multiple library lengths. Since $\rho$ estimates the predictability of one's behaviour given another person's behaviour, it can be treated as a measure of influence. Hence, we define influence score by $\rho$. If $\rho$ of $u_1$ is generally increasing with the library length, and its maximum is sufficiently great (with a threshold $\theta$), it indicates that it is possible to estimate $X_{u_1}$ from $X_{u_2}$. Therefore, in such case, we imply that the behaviour of $u_1$ drives $u_2$ (i.e., $u_1 \Rightarrow u_2$). It should also be noted that both $u_1 \Rightarrow u_2$ and $u_2 \Rightarrow u_1$ can happen at the same time. If $u_1 \Rightarrow u_2$, we mark both $u_1$ and $u_2$ to be suspected coordinated users. It is possible that $u_1$ is influenced by any other $u_3$ at different partitions of $X_{u_1}$. Even though it could hinder the variation of $\rho$, CCM successfully recovers from it since we embed only a part of history instead of the whole history. Specific hyperparameters and methodologies that are used in the submodules are given in Section \ref{subsec:experimental-setup}.

\subsubsection{Pairwise comparison.}

The computational expense associated with pairwise comparisons of $N$ users can be substantial ($^N{C}_2=\mathcal{O}(N^2)$), yet accurate. In response, we devise an optimization strategy based on the observation of our raw results in Section \ref{subsec:optimizations}.

\section{Experiments}
\label{sec:experiments}
\subsection{Data}
\label{subsec:data}

\begin{figure}
  \centering
  \includegraphics[width=\linewidth]{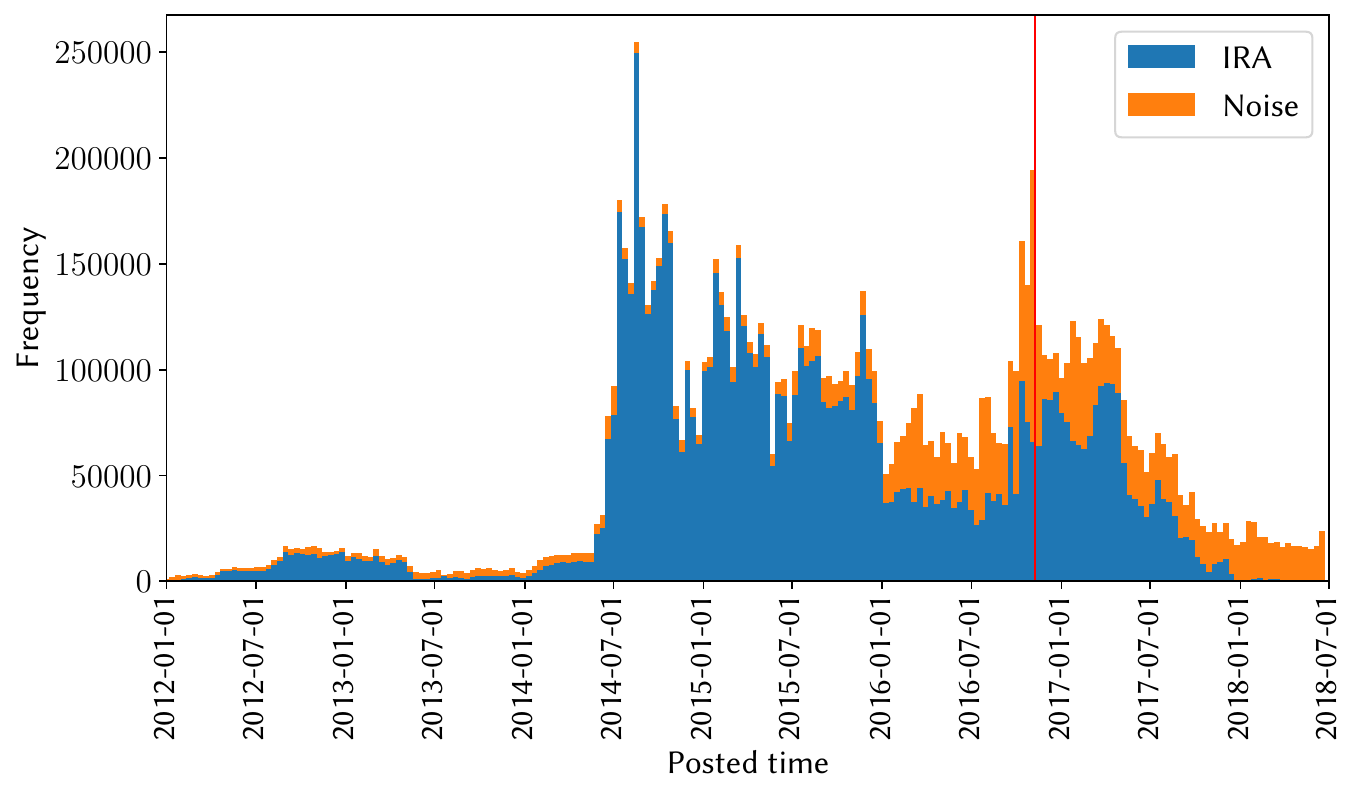}
  \caption{Stacked distribution of IRA activities and extracted noise tweets    across time. The bin size for the x-axis is 1 million seconds ({\raise.17ex\hbox{$\scriptstyle\mathtt{\sim}$}}11.6 days). The red vertical line shows the election date.}
  \label{fig:data-distribution}
\end{figure}

We experiment on the dataset of the activity of Russia's Internet Research Agency (IRA) influencing the 2016 USA presidential elections \citep{permanentselectcommitteeonintelligenceExposingRussiaEffort2018, roberts.muellerUNITEDSTATESAMERICA2018}, which consists of confirmed coordinated activities. This is a widely used dataset for detecting coordination \citep{weberAmplifyingInfluenceCoordinated2021,sharmaIdentifyingCoordinatedAccounts2021, zhangVigDetKnowledgeInformed2021,weberTemporalNuancesCoordination2022} due to the availability of ground truth. The dataset consists of 8.76 million tweets posted by 3613 users. The dataset originally consisted of the following fields; Tweet id, User id, User display name, User screen name, User reported location, User profile description, User profile url, Follower count, Following count, Account creation date, Account language, Tweet language, Tweet text, Tweet time, Tweet client name, Replied tweet id, Replied user id, Quoted tweet id, Whether the tweet is a retweet, Retweeted user id, Retweeted tweet id, Latitude where the tweet is posted, Longitude where the tweet is posted, Quote count, Reply count, Like count, Retweet count, List of hashtags, List of urls, List of user mentions, List of poll choices if the tweet includes a poll. Figure \ref{fig:data-distribution} shows the distribution of activity across the time.


In order to test the effectiveness of a coordination detection model, we introduce a set of noisy background events to the IRA dataset, since the IRA dataset only contains the set of coordinating users. For that purpose, we scraped Twitter data for that period of time which includes the same popular hashtags in the IRA dataset using the Twitter API v2 for academics. The criteria that were used to extract data were: posted time between 2008 and 2018, marked location anywhere in the USA, contains either one of the following hashtags - \emph{Election2016}, \emph{MAGA}, \emph{MakeAmericaGreatAgain}, \emph{AmericaFirst, DonaldTrump}, \emph{WakeUpUSA}, \emph{Trump}, \emph{TrumpTrain}, \emph{HilaryClinton, Trump2016}, \emph{DrainTheSwamp}, \emph{TrumpPence16}, \emph{tcot}, \emph{POTUS}, \emph{GOP}, \emph{Resist}, \emph{UniteBlue}, \emph{NeverHillary}, \emph{ElizabethWarren}, \emph{WeThePeople}, \emph{IllegalAliens}, \emph{TrumpRussia, ImWithHer}, \emph{GayHillary}, \emph{WakeUpAmerica}. The above set of hashtags were the top-occurring hashtags in the original IRA dataset to ensure that the noise data belongs to the same ongoing discussions at that period of time. The background data of normal users consists of 2.80 million tweets from 333,000 of users. The distribution of coordinating tweets and the noisy tweets are shown in Figure \ref{fig:data-distribution}. High activity is apparent near the election time period (November 2016).

\subsection{Experimental Setup}
\label{subsec:experimental-setup}

\definecolor{COrange}{HTML}{ff812c}
\definecolor{CPink}{HTML}{ff6ccd}
\definecolor{CGreen}{HTML}{63c227}
\definecolor{CBlue}{HTML}{58A5FF}
\definecolor{CBrown}{HTML}{766951}

\begin{figure*}[t!]
    \centering
    \subfloat[\centering \label{fig:results-without-nmf}]{{\includegraphics[width=0.32\linewidth]{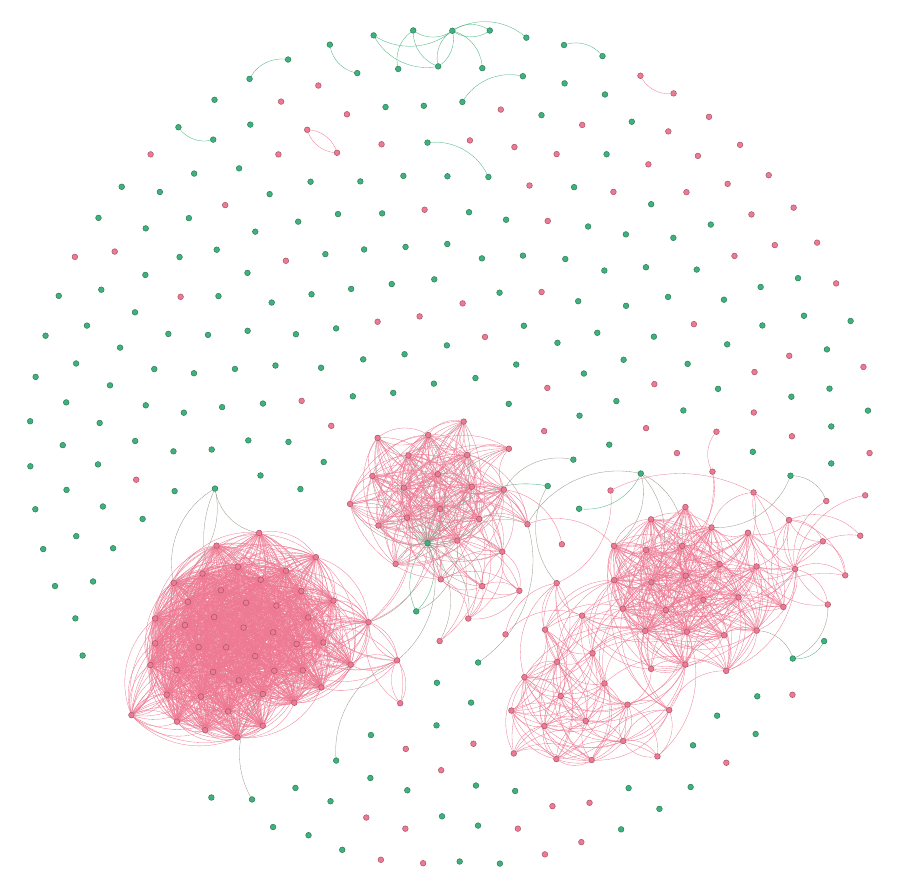} }}%
    \subfloat[\centering \label{fig:results-marked-communities}]{{\includegraphics[width=0.32\linewidth]{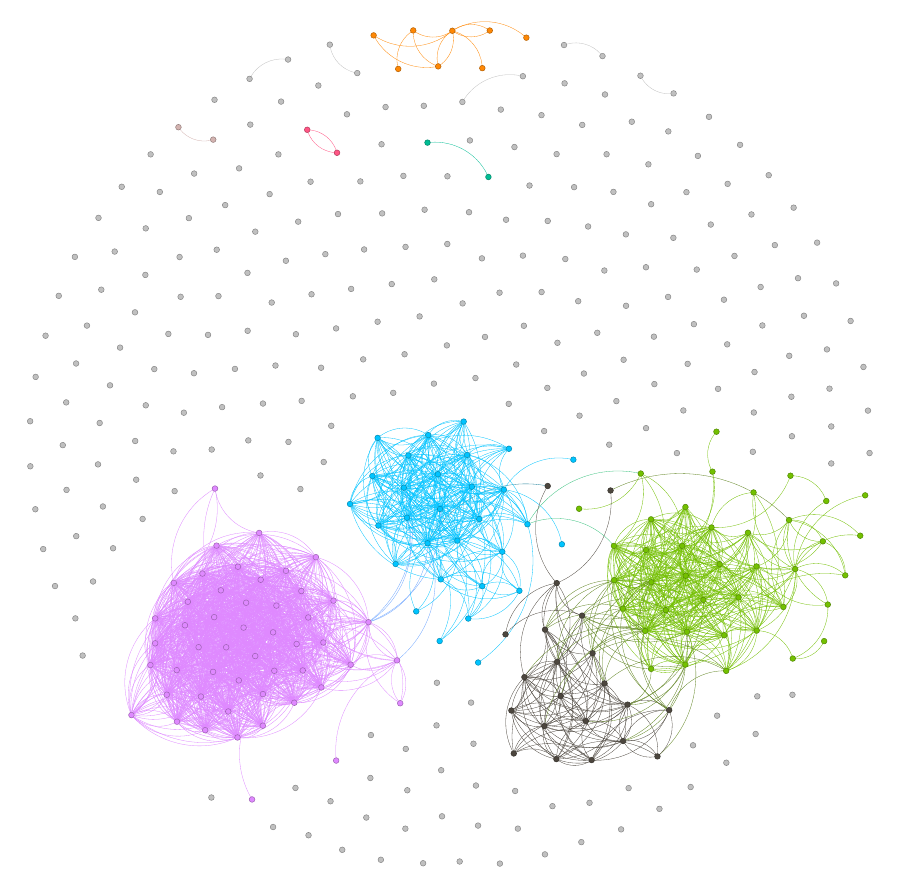} }}%
    \subfloat[\centering \label{fig:results-marked-by-topics}]{{\includegraphics[width=0.32\linewidth]{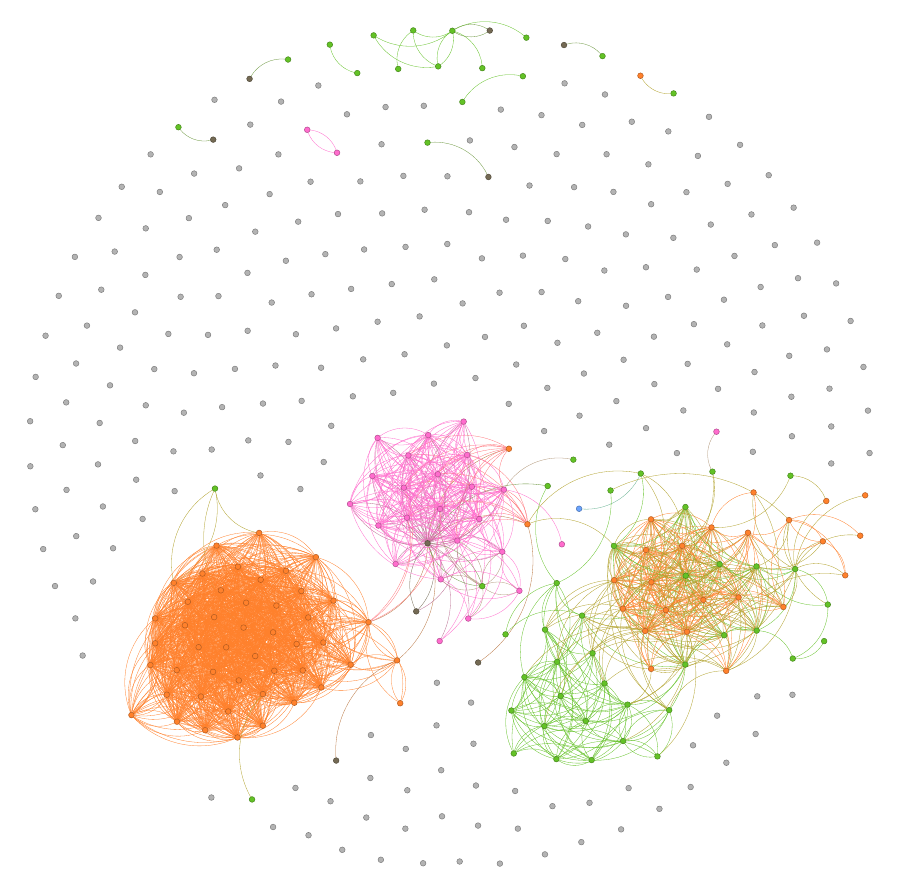} }}%
    \caption{Influence graphs for $N_C=200, N_N=200$. Each edge represents an edge identified by CCM. The edge color is simply an average color between the vertices. (a) Pink vertices are known coordinating users. Green vertices are known normal users. (b) Vertex color represents the community identified \cite{Blondel_2008}. (c) Vertex color represents the  topic of discussions of each user identified by CCM \textcolor{COrange}{\CIRCLE} -- General, \textcolor{CGreen}{\CIRCLE} -- Trump vs. Hillary, \textcolor{CPink}{\CIRCLE} -- News,  \textcolor{CBrown}{\CIRCLE} -- Democratic Party, \textcolor{CBlue}{\CIRCLE} -- Emotions.}%
    \label{fig:results-before-nmf}%
\end{figure*}
\subsubsection{Parameters.}

We found that a bin size $I$ of 60 minutes and a lag  $\tau$ of minimum value 1, and an embedding size $E$ as 10 to be parameter values that yield the best results. The threshold $\theta$ was chosen as 0.5. We split the time trace vectors into 3:1 ratio for train, test datasets.

\subsubsection{Time Intervals.}
The time period $(t_{\text{start}}, t_{\text{end}})$ was chosen such that it includes the election time period (November 2016) with the assumption that the coordinated activity was at a maximum during that period of time. Thus, $t_{\text{start}}$ was chosen as July 2016 and $t_{\text{end}}$ was chosen as November 2016.

\subsubsection{Extracting Top Users.}

For the tests to be fair, we mix the top $N_C$ number of known coordinating users and top $N_N$ number of known normal users from the above IRA dataset. The users for tests were selected based on the frequency of activities in the testing time period in order to ensure we have enough data to cross map each and every user.

\subsubsection{Submodules.}

To measure correlation $\rho$, we use Pearson's correlation method. In order to measure the general increase in correlation values for multiple library lengths, a straightforward linear regression was conducted, and the resulting gradient was used to assert the growth.

\subsection{Results}
\label{subsec:results}


Following is a report of our results for $N_C=200$ and $N_N=200$. Out of $^{400}{C}_2$ number of user pairs checked, 2404 pairs were identified as coordinating pairs. Out of such pairs,
\begin{itemize}
    \item 2319 (96.5\%) were known coordinating – coordinating pairs.
    \item 63 (2.6\%) were known coordinating – normal pairs.
    \item 22 (0.9\%) were known normal – normal pairs.
\end{itemize}
Since we mark each user who belongs to at least one $u_1 \Rightarrow u_2$ pair as coordinating, our model detected 165 users as coordinating. For that case, the precision is 80.0\% and recall is 72.0\% for detecting a coordinated user out of a mix of users. The model took 43 minutes to train and predict on an M1 MacBook. Refer to Table \ref{tab:runtimes} for other dataset sizes.

Figure \ref{fig:results-without-nmf} displays a graph we constructed using vertices as users and edges as influence flows identified by our model. It is apparent that there are four visible clusters of tightly coupled users for this sample. Figure \ref{fig:results-marked-communities} and \ref{fig:results-marked-by-topics} are described in the following subsections. There, we demonstrate how we exploited the clustered nature of users to optimize our model.

\subsection{Optimizations}
\label{subsec:optimizations}

\subsubsection{Motivation.}

In order to formally identify sub-communities in the graph in Figure \ref{fig:results-without-nmf}, we performed community detection \citep{Blondel_2008} on our results. The community detection algorithm could detect 5 main sub-communities. The colored sub-communities are shown in Figure \ref{fig:results-marked-communities}.

In a perfect scenario, say we could detect $n$ equal sized clusters in a set of users $U$ of size $N$. If we only compare user pairs within the clusters, our search space is reduced from $^{N}C_2$ to $n \times ^{N/n}C_2$. The relative decrease in run time is:

\begin{align*}
\frac{n \times ^{N/n}C_2}{^{N}C_2}=\frac{n\frac{N/n\left(N/n - 1\right)}{2}}{\frac{N(N - 1)}{2}}&=\frac{N-n}{n(N-1)}\\
&\sim \frac{1}{n}\text{, Given }N \gg n
\end{align*}

This is a huge increase in performance in the best case. A suitably engineered clustering technique could achieve nearly equal clusters and hence can achieve this much performance increase in terms of computational time to our model.

We experimented with different clustering techniques, and compared those results with the sub-communities identified above as the baseline. For a comparison metric, Adjusted Rand Score \cite{Steinley2004} was used. Since we detected 5 sub-communities using community detection for the above sample, for comparison, we used $n=5$ as the number of components (clusters) for each clustering method, since we observed 4 large visible clusters and a small cluster at the top of the graph in Figure \ref{fig:results-marked-communities}. Table \ref{tab:different-topic-models} shows that NMF (Non-negative Matrix Factorization) topic modelling yields the best results out of the tested methods.

\begin{table}
    \caption{Comparing different clustering techniques with the identified communities.}
    \label{tab:different-topic-models}
    \centering
    \begin{tabular}{lc}
        \toprule
    Method                 & Adjusted Rand Score \\
    \midrule
    Baseline (Communities) & 1                   \\
    \textbf{NMF}                    & \textbf{0.38}                \\
    K-Means                & 0.09                \\
    DBSCAN                 & 0.11                \\
    OPTICS                 & 0.12                \\
    Feature Agglomeration  & 0.21              \\
    \bottomrule 
    \end{tabular}
    \end{table}

\begin{figure*}[t!]
    \centering
    \subfloat[\centering \label{fig:results-after-topics}]{{\includegraphics[width=0.3\linewidth, trim=0 1.5cm 0 1.5cm, clip]{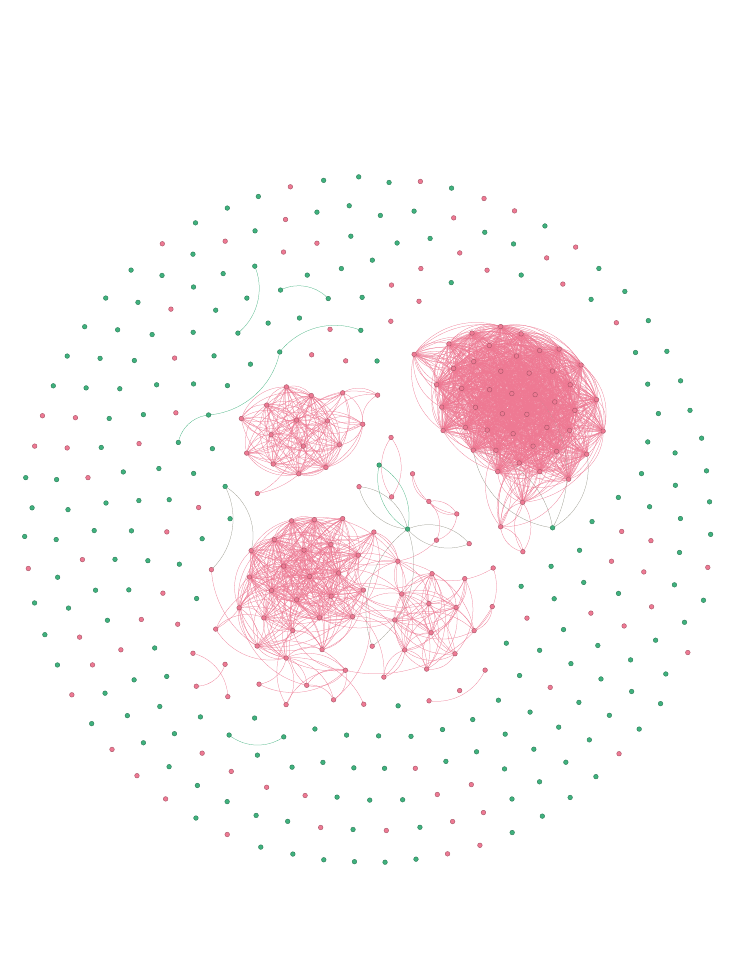} }}%
    \subfloat[\centering \label{fig:results-after-topics-with-topics}]{{\includegraphics[width=0.3\linewidth, trim=0 1.5cm 0 1.5cm, clip]{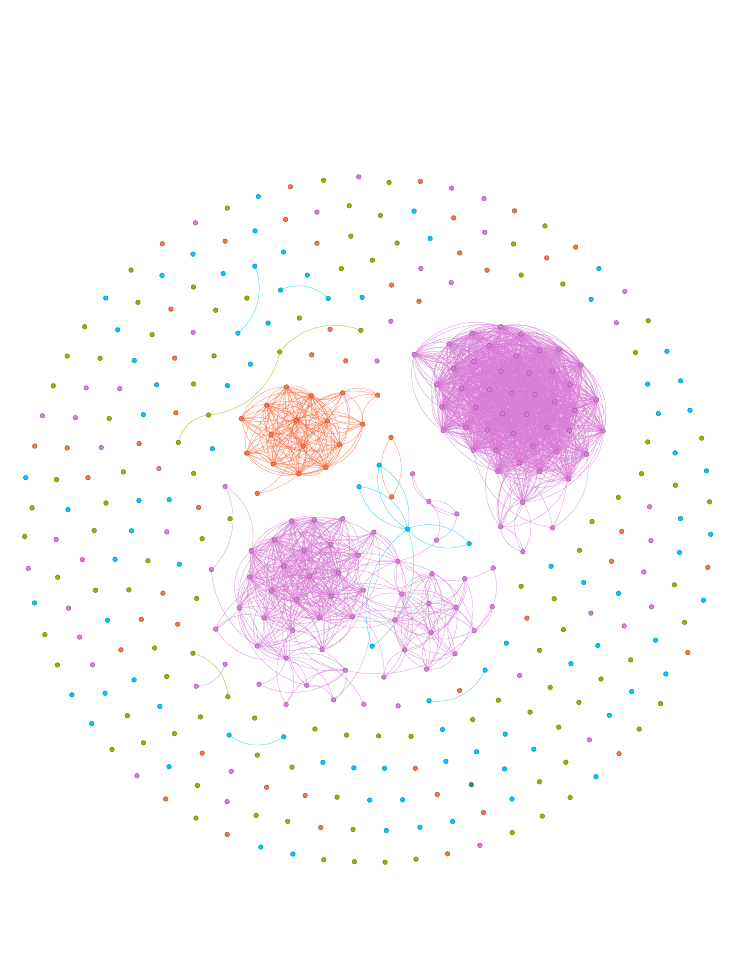} }}%
    \caption{Influence graph for $N_C=200, N_N=200$. Each edge represents an edge identified by CCM after isolating user groups by topics. The edge color is simply an average color between the vertices. (a) Pink vertices are known coordinating users. Green vertices are known normal users. (b) Vertex color represents the  topic of discussions of each user. \textcolor{COrange}{\CIRCLE} -- News, \textcolor{CPink}{\CIRCLE} -- General, \textcolor{CBlue}{\CIRCLE} -- Democratic Party.}%
    \label{fig:results-after-topics-all}%
\end{figure*}

\begin{figure*}[t!]
    \centering
    \subfloat[\centering \label{fig:results-after-topics-400}]{{\includegraphics[width=0.3\linewidth, trim=0 1.5cm 0 1.5cm, clip]{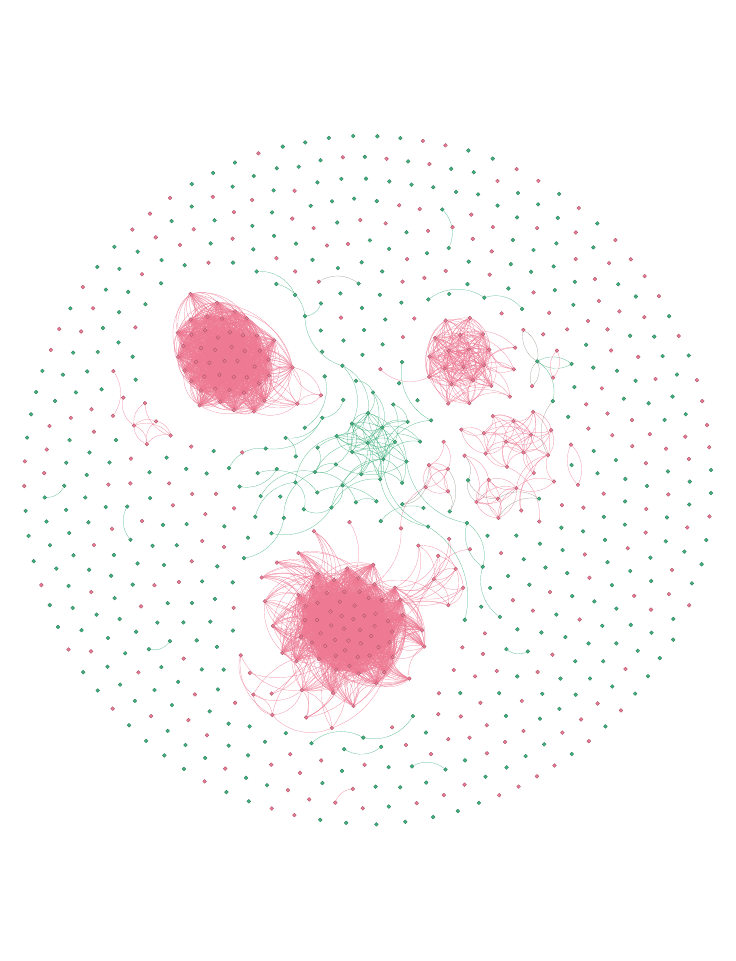} }}%
    \subfloat[\centering \label{fig:results-after-topics-400-topics}]{{\includegraphics[width=0.3\linewidth, trim=0 1.5cm 0 1.5cm, clip]{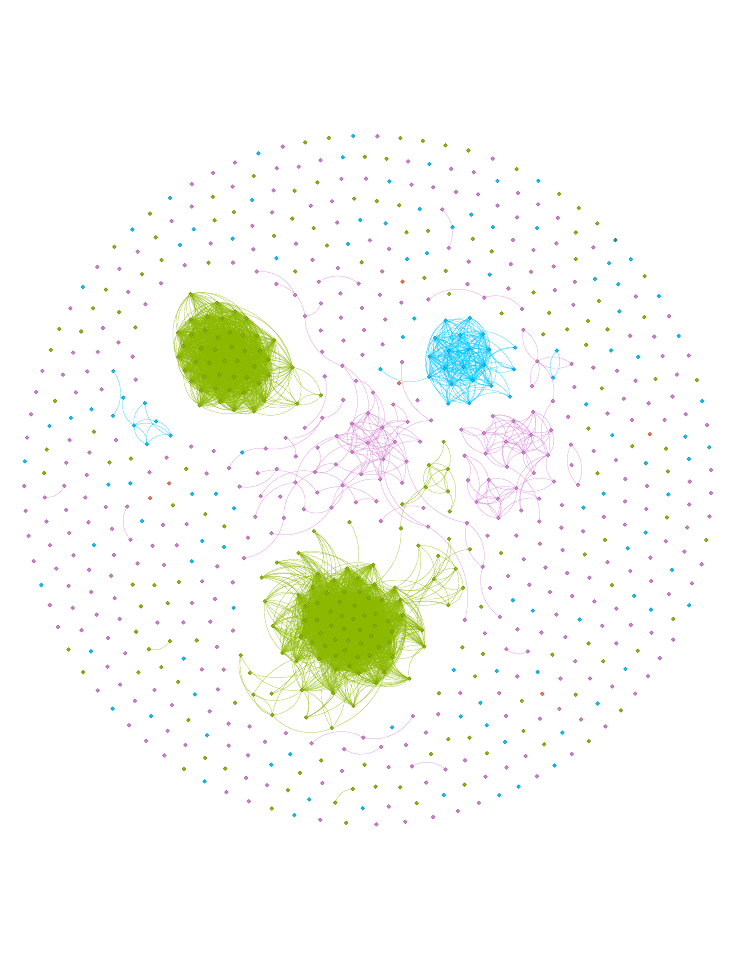} }}%
    \caption{Influence graph for $N_C=400, N_N=400$. Each edge represents an edge identified by CCM after isolating user groups by topics. The edge color is simply an average color between the vertices. (a) Pink vertices are known coordinating users. Green vertices are known normal users. (b) Vertex color represents the  topic of discussions of each user. \textcolor{CGreen}{\CIRCLE} -- General, \textcolor{CPink}{\CIRCLE} -- Politics, \textcolor{CBlue}{\CIRCLE} -- News.}%
    \label{fig:results-after-topics-400-all}%
\end{figure*}

\begin{table*}[t!]

    \caption{Comparing original results and results with topic clustering including runtimes in minutes. CC -- the number of known coordinating - coordinating pairs detected by the model, CN -- the number of known coordinating - normal pairs detected by the model, and NN -- the number of normal - normal pairs detected by the model.}
    \label{tab:runtimes}
    \centering
\begin{tabular}{llccccccc}
    \toprule
    Dataset & Method & Runtime & CC & CN & NN & Precision & Recall & F1 Score \\

    \midrule
    \multirow{2}{*}{$N_C=100, N_N=100$} & CCM & 11 & 610 & 41 & 10 & 87.3\% & \textbf{62.0\%} & 72.5\%     \\
                                        & CCM + NMF & 4.2 & 524 & 7 & 7 & \textbf{91.0\%} & 61.0\% & \textbf{73.0\%}   \\
    \cmidrule(lr){2-9}
    \multirow{2}{*}{$N_C=200, N_N=200$} & CCM & 43 & 2319 & 63 & 22 & 80.0\% & \textbf{66.0\%}   & 72.0\%     \\
                                        & CCM + NMF & 11.4 & 1818 & 12 & 7 & \textbf{91.4\%}   & 64.0\% & \textbf{75.3\%}   \\
    \cmidrule(lr){2-9}
    \multirow{2}{*}{$N_C=400, N_N=400$} & CCM & 164 & 4453 & 123 & 153 & 66.1\% & \textbf{52.2\%} & 58.4\%       \\                                                       
                                        & CCM + NMF & 60.5 & 4106 & 13 & 119 & \textbf{69.8\%}    & 49.8\% & \textbf{58.1\%} \\
      \bottomrule
\end{tabular}
\end{table*}

\subsubsection{Topic modelling.}

NMF \cite{NMF} is a matrix factorization technique that decomposes a non-negative $W\times H$ sized matrix into two matrices of size $W \times n$ and $n \times H$ as a product. $n$ is a significantly smaller number than $W$ and $H$. Due to the clustering property of NMF, semantically related terms are automatically grouped, forming distinct topics. In order to perform NMF, a document term matrix is constructed while TF-IDF weight adjustment is applied to the dataset to ensure term importance. Given an $n$, NMF decomposes this matrix into two matrices: (1) Document term matrix ($W \times n$) - Each row represents a document, and each column represents a topic, indicating the document's distribution over topics. (2) Term-topic matrix ($n \times H$) - Each row represents a topic, and each column represents a term, indicating the importance of each term within each topic. The challenge here is to find the least number of topics that partitions the dataset into semantically different subsets. Practically, maximizing the Average Silhouette Score \cite{ROUSSEEUW198753} can be recommended to determine the number of clusters $n$.

\subsubsection{Observations and optimization methodology.}

Each tweet was treated as a document. Both English and Russian stop-words were removed and the documents were vectorized using TF-IDF vectorization. Then, NMF was applied to the matrix constructed by concatenating the TF-IDF vectors. For the following samples, the number of topics was chosen as 5 to run NMF due to the observations made in Figure \ref{fig:results-marked-communities}. To derive the cluster of each of user they belongs to, each tweet of a user is concatenated into a single document. Then, the trained NMF was used to predict the topic to which that long document belongs. The percentage shows the proportion of the number of people who belong to each topic out of everyone who was tested. The following are the top words that appeared in topics along with our own interpretation of the topic in a single word/phrase. 

\begin{itemize}
    \item need, make, think, life, want, know, people, just, like, don -- General (44.2\%)
    \item campaign, debate, cnn, says, vote, politics, donald, clinton, hillary, trump -- Trump vs. Hillary (33.9\%)
    \item killed, new, state, cbs, man, says, kansas, police, world, news -- News (15.8\%)
    \item far, muslim, isis, president, american, hillary, america, usa, obama, tcot -- Democratic Party (5.5\%)
    \item ll, let, heart, oh, fall, hate, song, true, life, love -- Emotions (0.6\%)
\end{itemize}

Figure \ref{fig:results-marked-by-topics} shows a graph of the users colored by the topic they are associated with. It is apparent that some topics clearly overlap with the clusters we identified using community detection in Figure \ref{fig:results-marked-communities}.

For NMF to be used as an optimization step, we cluster users using the topic. Then, we do pairwise cross mapping for each user pair inside the cluster. We evaluated the performance of our CCM model while exploring the impact of incorporating NMF on accuracy and runtime for different sizes of datasets. Table \ref{tab:runtimes} summarizes the results. CCM combined with NMF demonstrated higher precision than CCM alone, identifying a greater proportion of true coordinating pairs among those detected. Recall remained relatively consistent across both methods due to the reduction in search space, suggesting similar abilities to detect existing coordinated pairs. CCM + NMF consistently exhibited faster runtimes compared to CCM alone. This suggests that topic clustering can significantly improve efficiency without compromising accuracy. 

CCM + NMF results in lower precision and recall for the $N_C=400, N_N=400$ case. This decline is largely attributed to the selection of top active users, as inadequate data points hinder CCM's ability to identify relevant users. Nonetheless, if there is sufficient data and the coordination actions are well preserved in the time series data, our model would not suffer from these issues since it compares users pairwise, making the calculation independent of the dataset size. Therefore, the amount of data points and the preservation of coordination in time series are important for scalability.

Figure \ref{fig:results-after-topics-all} shows the derived influence graphs with this optimization for $N_C=200$ and $N_N=200$. Figure \ref{fig:results-after-topics-400-all} shows results for $N_C=400$ and $N_N=400$.

\subsubsection{Receiver operating characteristic (ROC curve) after topic clustering.}

\begin{figure}[t!]
    \centering
    \includegraphics[width=\linewidth]{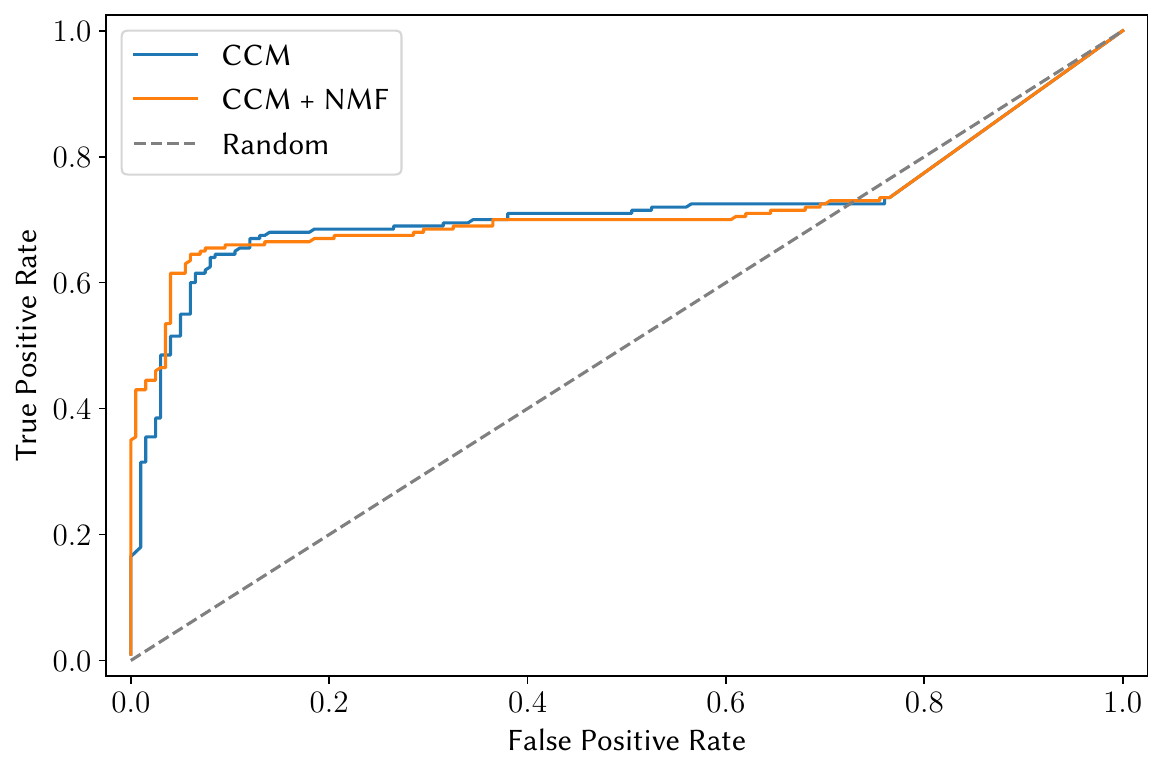}
    \caption{ROC curves of CCM and CCM + NMF models for $N_C=200, N_N=200$ case at different threshold values.}
    \label{fig:auc-roc}
  \end{figure}

The corresponding ROC curve for the CCM + NMF model for $N_C=200, N_N=200$ case reveals that the performance of the refined approach closely aligns with that of the CCM model for the same dataset (Figure \ref{fig:auc-roc}). In detail, the area under the curve (AUC) for CCM and CCM + NMF is 0.7219 and 0.7221 respectively. This observation is an important indicator that the reduction in search space does not lead to a noticeable degradation in performance. Consequently, CCM + NMF emerges as a viable alternative, offering a more efficient yet equally effective solution for identifying coordination. Furthermore, the optimal thresholds for maximizing Youden's J statistic \cite{youden50} were identified at 0.56 for the CCM model and 0.49 for the CCM + NMF model, which supports the selection of 0.5 as a reasonable threshold for these analyses.

  \begin{figure}[t!]
      \centering
      \includegraphics[width=\linewidth]{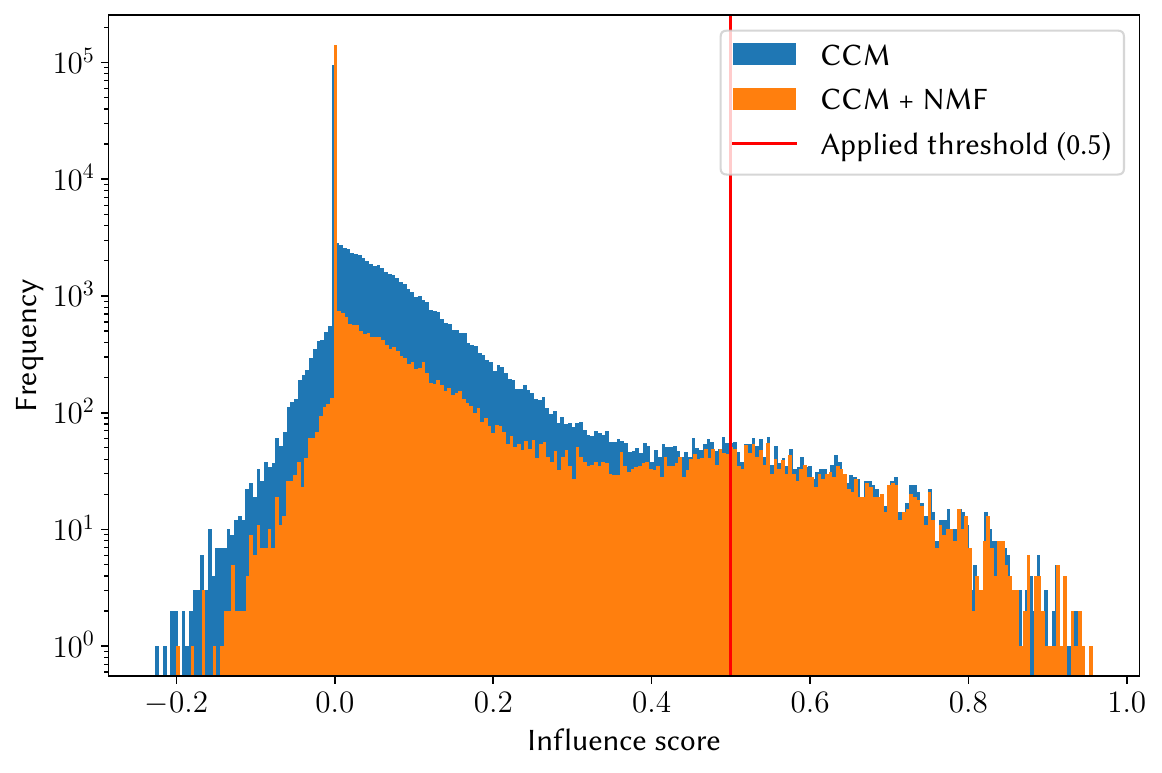}
      \caption{The distribution of influence scores ($\rho$) yielded by CCM and CCM + NMF models for $N_C=200, N_N=200$ case. Bin size is approximately 0.005.}
      \label{fig:influence-distribution}
    \end{figure}

\subsubsection{The distribution of influence scores.}
The distribution of influence scores ($\rho$) yielded by both CCM and CCM + NMF is plotted to illustrate the capability of NMF in detecting less significant links (Figure \ref{fig:influence-distribution}). The log scaled y-axis emphasizes the difference in the frequency distributions below the threshold. Simultaneously, the negligible difference in the frequency distributions above the threshold highlights NMF's ability to preserve critical links. This clearly highlights the discriminative power of NMF.

\subsection{Baseline comparisons}
\label{subsec:baseline-comparisons}

We choose the following baselines to compare our results.

\begin{enumerate}
    \item \emph{GC \citep{granger69}.} Granger Causality (GC) is a method that incorporates statistical testing to determine the causality between two time series. Given two users, we use a $p$-value of 5\% to determine if the two activity traces are causally related. This method was used in place of CCM in our model (without NMF) to determine the effectiveness of CCM in identifying coordinated users.
    \item \emph{LCN + HCC \citep{weberAmplifyingInfluenceCoordinated2021}.} This approach aims to identify coordinated communities using community detection on user similarity graphs. The temporal aspect is considered by a windowing mechanism. We set the window length parameter to 10 days.
    \item \emph{Tweet language.} Since most (82\%) of the data in the coordinated set of users are in Russian and most (93\%) of the data in the noise data are in English, we compare our results with the results of a naive model that only uses the language to determine the coordination status. This model simply classifies a user to be coordinated if the language is Russian.
    \item \emph{AMDN-HAGE.} \citep{sharmaIdentifyingCoordinatedAccounts2021} This is the SOTA for identifying coordinated users. We use the same set of hyperparameters except the threshold to determine the output influence values. Instead, we maximize the F1 score to determine it.
\end{enumerate}

Table \ref{tab:compare} showcases comparisons of our results with the above baselines. In all cases, CCM outperforms GC in terms of F1 score. Specifically, for $N_C=200, N_N=200$ case, GC detects 77191 edges out of 79800 possible bidirectional edges leading to classifying all users in the sample as coordinating users. This leads to a recall of 100\%. With GC, out of the 77191 identified edges, 19705 (25.5\%) were known coordinating-coordinating pairs. These statistics can be compared with Table \ref{tab:runtimes}. Thus, the choice of CCM for determining causality between time series data is justified. The CCM model achieved the highest precision, indicating a superior ability to accurately identify true coordinating pairs among those detected. This suggests CCM's effectiveness in minimizing false positives, a crucial aspect of coordinated user detection. AMDN-HAGE exhibited the highest recall, suggesting its strength in detecting the majority of existing coordinating users. However, its relatively low precision indicates a higher propensity for false positives, perhaps due to the limited timeframe of the dataset. CCM + NMF achieved the highest F1 scores for datasets with 200 and 400 users, demonstrating a favorable balance between precision and recall. This highlights its potential to provide more comprehensive and accurate coordination detection compared to the other baselines. However, its performance for 800 users was lower compared to LCN + HCC, indicating potential room for further optimization.

\begin{table*}[t!]
    \caption{Results for detecting coordinated users using different methods. $N_C$ - number of coordinating users in the dataset, $N_N$ - number of normal users in the dataset.}
    \label{tab:compare}
    \centering
    \begin{tabular}{lccccccccc}
      \toprule
      \multirow{2}{*}{Method} & \multicolumn{3}{c}{$N_C=100, N_N=100$} & \multicolumn{3}{c}{$N_C=200, N_N=200$} & \multicolumn{3}{c}{$N_C=400, N_N=400$} \\
      \cmidrule(lr){2-4} \cmidrule(lr){5-7} \cmidrule(l){8-10} 
      & Precision & Recall & F1 Score & Precision & Recall & F1 Score & Precision & Recall & F1 Score \\
      \midrule
      GC & 48.0\% & 98.0\% & 64.4\% & 50.0\% & 100\% & 66.7\% & 45.0\% & 80.0\% & 57.6\% \\
      Tweet language & 64.0\% & 80.0\% & 71.1\% & 66.0\% & 81.0\% & 72.7\% & 66.0\% & 75.0\% & 70.2\% \\
      LCN + HCC & 76.1\% & 63.0\% & 68.9\% & 77.3\% & 65.0\% & 70.6\% & \textbf{81.5\%} & 70.4\% & \textbf{75.5\%} \\
      AMDN-HAGE & 50.0\% & \textbf{98.0\%} & 66.2\% & 50.4\% & \textbf{100\%} & 67.0\% & 50.6\% & \textbf{100\%} & 67.2\% \\
      \midrule
      CCM & 87.3\% & 62.0\% & 72.5\% & 80.0\% & 66.0\% & 72.0\% & 66.1\% & 52.2\% & 58.4\% \\
      CCM + NMF & \textbf{91.0\%} & 61.0\% & \textbf{73.1\%} & \textbf{91.4\%} & 64.0\% & \textbf{75.3\%} & 70.1\% & 49.3\% & 57.9\% \\
      \bottomrule
    \end{tabular}
  \end{table*}

\subsection{Leader-follower behaviour}
\label{subsec:leader-follower}

Recall that influence is a directional relationship between users. A leader on an OSN could be someone who originates content or significantly contributes to the spread of content, ideas or trends in the network. Such leaders can be identified by examining how often they are retweeted/mentioned, having high degree centrality in the  influence graph. On the other hand, a follower is someone who consumes or amplifies the content of leaders. Vertices whose indegree is high but outdegree is relatively low in the influence graph could be a user with a follower personality.

Define \emph{net-degree} to be the difference between the outdegree($\text{deg}^+(v)$) and indegree($\text{deg}^-(v)$) i.e., $\text{ndeg}(v)=\text{deg}^+(v) - \text{deg}^-(v)$. We inspected the influence graphs and checked the users who have the top net-degrees. To verify our results, for each user in the sample, we listed the number of times they were retweeted and the number of times they were mentioned. We recorded the percentile they belong in both categories. Table \ref{tab:leader-follower} demonstrates our results. The following are the user display names associated with the top users and some details about them (Russian names are translated to English).

\begin{itemize}
    \item $u_1$: Open Russia -- Open Russia constitutes a political organization established by the exiled Russian businessman Mikhail Khodorkovsky \cite{khodorkovskyMikhailKhodorkovskys}.
    \item $u_2$: $<$Anonymized$>$ -- Conservative political science commentator (according to their profile description on Twitter).
    \item $u_3$: Ramzan Kadyrov -- A Russian politician, currently the head of the Chechen Republic
    \item $u_4$: Moscow Bulletin -- A bulletin service
    \item $u_5$: Bulletin of Novosibirsk -- A bulletin service
\end{itemize}

According to the above information, the influence graph combined with net-degree sorting was able to pick important users without prior knowledge of the content they post, thus supporting the reliability and effectiveness of the CCM methodology.

\begin{table*}[t!]

    \caption{Users with top net-degree in the derived influence graph without optimizations. The number of tweets, retweets and mentions are calculated within the sample time window of 4 months from July 2016 with $N_C=200, N_N=200$. User-ids are hidden due to Twitter terms of service.}
    \label{tab:leader-follower}
    \centering
\begin{tabular}{lccC{2cm} cc}
      \toprule
      User & IRA user? & Number of tweets & Net-degree (Indegree, Outdegree) & Retweets (Percentile) & Mentions (Percentile) \\
      \midrule
      $u_1$ & Yes & 788 & 14 (8, 22) & 34 (94.8\%) & 36 (93.8\%) \\
      $u_2$ & No & 6729 & 8 (20, 28) & 88 (95.8\%) & 112 (96.5\%) \\
      $u_3$ & Yes & 1061 & 7 (14, 21) & 89 (96.0\%) & 91 (95.5\%) \\
      $u_4$ & Yes & 1257 & 6 (19, 25) & 3 (75.1\%) & 5 (74.5\%) \\
      $u_5$ & Yes & 891 & 5 (14, 19) & 6 (81.3\%) & 6 (77.2\%) \\

      \bottomrule
\end{tabular}
\end{table*}

\subsection{Uncovering coordinated behaviours}
\label{subsec:uncovering}

\begin{figure}[t!]
    \centering
    \includegraphics[width=\linewidth]{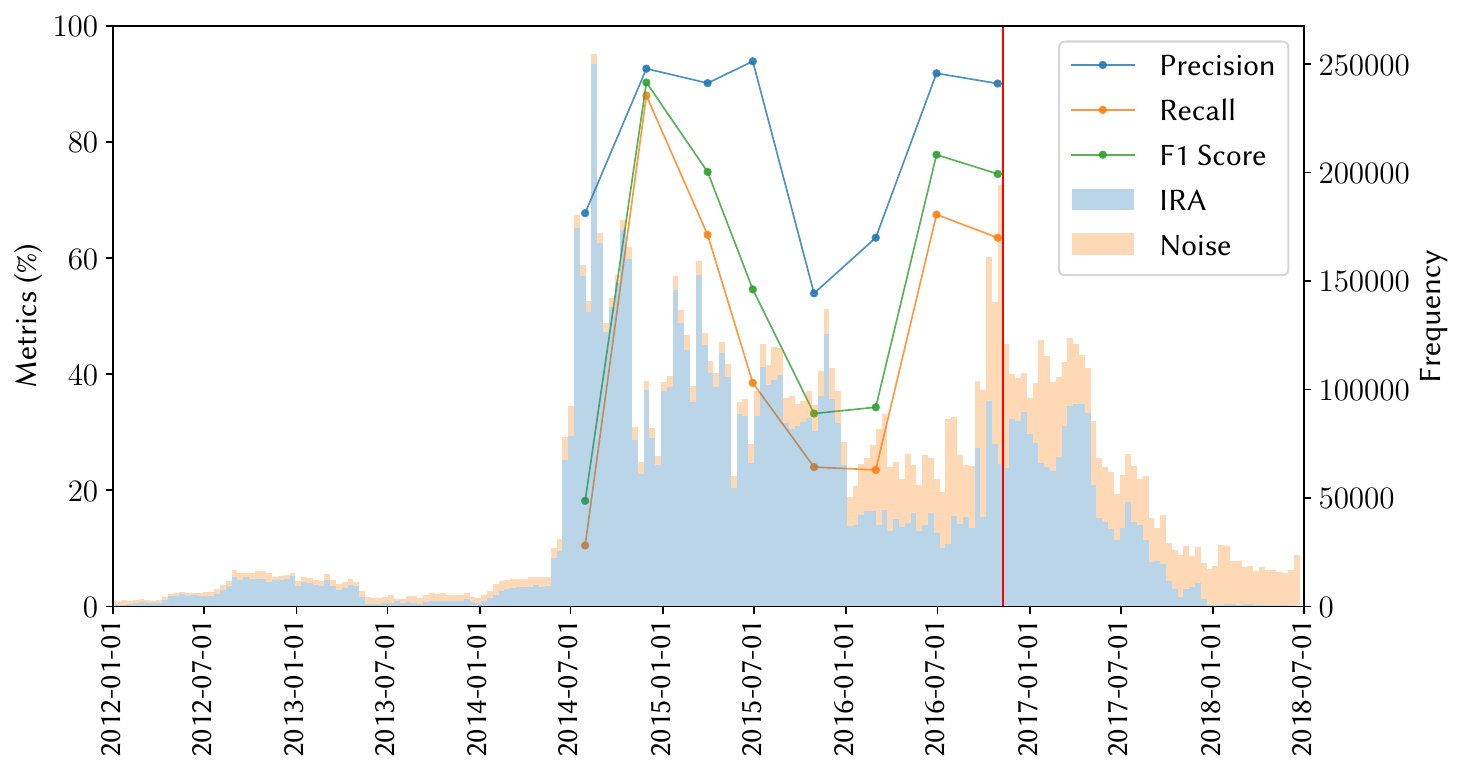}
    \caption{Results at different times for the same set of users. This indicates that the IRA Twitter attackers were performing coordinated attacks even before 2016 US elections.}
    \label{fig:different-times}
  \end{figure}

We applied our model to the same set of users but to different periods of time. Interestingly, we get better results between November 2014 to July 2015 (See Figure \ref{fig:different-times}) compared to 2016 election times, which indicates higher coordination at that time. Upon inspection, we could observe that almost all the discussions were in Russian and they are related to mostly Russian and Ukrainian politics. There existed minor discussions related to US politics as well. Our results show that the IRA has been politically influencing different parts of the world even before 2016 US Elections even though the dataset was released due to their anomalous activity in 2016. The following are the top words translated from Russian to English that appeared in the identified topics in relevant time periods.

\begin{itemize}
    \item 4 months since November 2014
        \begin{itemize}
            \item politics, Vladimir, news, sanctions, rf, anti-sanctions, stoptank, Putin, Russia, EU
            \item will happen, prodigal, politics, Kiev tell the truth, defeat, plan, provocation of Kiev, Poroshenko, news, Ukraine
            \item next, situation, difference, interesting, battle of oligarchs, happening, provocation of Kiev, Kiev tell the truth, Kievsbilboing, Ukraine
            \item politics, world, read, interesting, retweet against Obama, Obama, politics, Obama, American plague, usa
            \item read, ready, looks like, interesting, battle of the oligarchs, provocation of Kiev, Kievsbilboing, Kiev tell the truth, gas sector, Ukraine
        \end{itemize}
    \item 4 months since March 2015
        \begin{itemize}
            \item EU, battle of the oligarchs, putin, alien, worthy, rf, quot, opinion, ukraine, news
            \item sanctions, rf, politics, politics, read, interesting, Obama, Ukraine, return California, USA
            \item foreign ministry, Poroshenko, sanctions, politics, coming, Klimkin, not easy, negotiations, Ukraine, Russia
            \item zelenskyrun, god, national, idea, Russia, Russians, read, written, interesting, Russian spirit
        \end{itemize}
\end{itemize}

\section{Case Study}
\label{sec:case-study}
\subsection{COVID-19 Dataset}
The development and validation of models aimed at detecting coordinated activities on social media platforms face significant obstacles, primarily due to the absence of datasets that are both representative of such activities and annotated with verifiable ground truth. This scarcity of labeled data complicates efforts to rigorously evaluate the performance of proposed detection methodologies.

In order to verify the performance of our model, we conducted a case study on the Twitter dataset related to COVID-19 curated by Lamsal \cite{Lamsal2020DesignAA}. The dataset spans one year and five days starting from 19 March 2020. It consists of approximately 75.9 million tweets generated by 10.8 million users. The tweets in this dataset were labelled with their stance in regard to anti-vax hesitancy (\emph{favor}, \emph{against}, and \emph{none}) in \citet{Zaidi2022TopicsIA}'s study using a stance detection tool based on OpenAI’s GPT transformer model \cite{Radford2018ImprovingLU}. 


During the COVID-19 pandemic, numerous bots were set up to collect and disseminate information about vaccines and public health measures. These bots were highly active, often retweeting content from mainstream media and official health sources to promote health protection and disseminate breaking news about COVID-19. The high level of public interest in the pandemic led to extensive resharing of these posts, creating a collective activity that appeared as coordinated behaviour \cite{info11100461}. This context underscores the relevance of the dataset for analysing coordinated actions, as it mirrors real-world scenarios where bot activity and public engagement form discernible patterns of coordination.

\subsection{Experimental Setup}

\subsubsection{Selecting a subset.}
In order to ensure that there is enough data to apply CCM, a sufficiently large subset of users from the dataset was selected. Thus, the most active 800 users were selected in the timespan starting from 1 December 2020 to the end of the dataset (nearly 5 months). This timespan was chosen since the first doses of COVID-19 vaccines were available since December 2020, a pivotal moment likely to influence user behaviour and potential coordination efforts.

\subsubsection{Parameters.}
We used a bin size $I=60 \text{ minutes}$ and a lag $\tau=1$, and an embedding size $E=16$ as parameters of CCM.

\subsection{Labelling users and user-user pairs}

The users were labelled by the content of their posts. User-user pairs were labelled by their semantic-agreeability.

\subsubsection{Stance.}
The tweets in the dataset were labelled by the stance. To assign a predominant stance to each user, we analysed the most frequent stance expressed across their tweets. It could be observed that, out of the 800 selected users, our analysis revealed the following distribution: 66.8\% exhibited an against stance (pro-vax), 36.1\% held an favor stance (anti-vax), and 24.1\% did not express a clear stance. Additionally, user-user pairs were defined as \emph{stance-agreed} if both individuals shared the same stance (either pro-vaccine or anti-vaccine); otherwise, they were considered \emph{stance-disagreed}.

\subsubsection{Topic.}
To further describe the semantic content of user posts, we applied NMF \cite{NMF}. We set the number of topics to six (by maximizing the topic coherence \cite{Rder2015ExploringTS}). In order to label a user with a topic, we concatenated all the tweets authored by that user and predicted the topic cluster using the trained NMF model. The following are the top words that appeared in topics along with our own interpretation of the topic in a single word/phrase. The percentage shows the proportion of users who belong to that topic cluster.

\begin{itemize}
    \item immunity, just, know, like, virus, time, need, don, people, vaccines    -- Immunity (60.8\%)
    \item pfizer, vaccinations, india, news, health, says, vaccination, vaccine, 19, covid -- Covid 19 Vaccination (21.1\%)
    \item today, health, covidvaccine, global, india, protests, pandemic, vaccine, vaccination, covid19    -- Global vaccine outreach (9.65\%)
    \item oxford, million, uk, says, astrazeneca, news, doses, pfizer, vaccine, coronavirus -- UK vaccines (8.15\%)
    \item quit, sex, alcohol, smoking, porn, extramarital, premarital, virus, corona, make -- Lifestyle (0.25\%)
    \item caring, pkpp, momennegaraku, prayformalaysia, jabatanpenerangan, watlakerdoh, kitajagakita, pmrdungun, kitateguhkitamenang, tidakpastijangankongsi
    -- Covid 19 in Malaysia (0.13\%)
\end{itemize}

Similar to stance-agreeability, if two users discuss similar topics, we define that they are \emph{topic-agreed}. Otherwise, we define that the two users are \emph{topic-disagreed}.

\subsection{Results}

This subsection details the findings of our model's performance in identifying semantic agreement among user pairs and distinguishing leader-follower dynamics within the dataset. It is important to clarify that the identification of an influence in the form $u_1 \Rightarrow u_2$ signifies the selection of the unordered pair $(u_1, u_2)$ as an identified pair. Consequently, the detection of a bidirectional influence is not requisite for the inclusion of a pair in the set of identified pairs.

\subsubsection{Proportion of semantically-agreed user pairs.}
Out of $^{800}{C}_2$ pairs of users, our model identified 1515 pairs of users. Out of the identified pairs, 965 (63.7\%) pairs were stance-agreed and 947 (62.5\%) pairs were topic-agreed. We could observe that the proportion of stance-agreed pairs and the topic-agreed pairs in the set of pairs identified by our model is increasing with thresholds higher than 0.5. This suggests that our model is capable in capturing influence flows among users who share semantic alignments, reinforcing its utility in detecting coordinated behaviour based on shared content and viewpoints.

\subsubsection{Leader-follower behaviour.}
We performed a study similar to Section \ref{subsec:leader-follower} on this dataset to further verify our model's ability to differentiate the nodes that are most influential and the nodes that are most influenced. The influence graph generated by our model was then inspected. The nodes that had the highest net-degree were selected as the most influential users and the nodes that had the lowest (negative) net-degrees were selected as the users who were most influenced by other users in the sample. 

The top five most influential nodes identified by our model are listed below, along with detailed descriptions of their accounts. Their in-degree, out-degree, stance, and the labelled topic are included in the parentheses.

\begin{itemize}
    \item All 435 Reps -- retweeting tweets by all members of the US House of Representatives ($\text{deg}^-(v)=23, \text{deg}^+(v)=107$, stance $=$ against (pro-vax), topic $=$ Immunity)
    \item Pub Health Monitor -- retweeting public health related tweets ($\text{deg}^-(v)=15, \text{deg}^+(v)=56$, stance $=$ against (pro-vax), topic $=$ Global vaccine outreach)
    \item all100Senators -- retweeting tweets by all members of the US Senate ($\text{deg}^-(v)=1, \text{deg}^+(v)=31$, stance $=$ against (pro-vax), topic $=$ Immunity)
    \item Devdiscourse -- retweeting international development news ($\text{deg}^-(v)=2, \text{deg}^+(v)=30$, stance $=$ against (pro-vax), topic $=$ Covid-19 vaccination)
    \item $<$Anonymized$>$ -- a researcher at a public health university ($\text{deg}^-(v)=2, \text{deg}^+(v)=30$, stance $=$ against (pro-vax), topic $=$ Immunity)
\end{itemize}

\begin{figure}[t!]
    \centering
    \includegraphics[width=\linewidth]{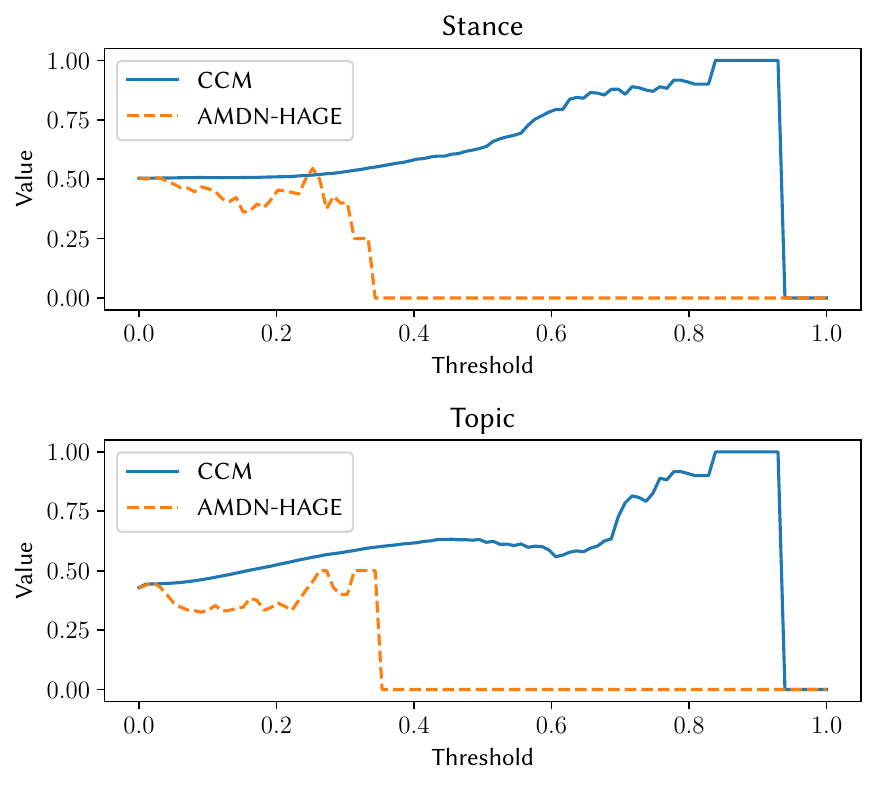}
    \caption{The variation of the proportion of semantically agreed pairs of users out of all the identified pairs of users under the variation of the threshold for both CCM and AMDN-HAGE models with respect to stance-agreeability (top) and topic-agreeability (bottom).}
    \label{fig:prec-recall-stance-topic}
  \end{figure}

Four of the above accounts can arguably be classified as bot accounts since their purpose is to retweet pre-determined accounts. However, apparently, the number of users whose behaviour is triggered by these few accounts are comparatively high. The users who are influenced by the most other users according to the influence graph are user accounts that are owned by normal people. These results demonstrate the applicability of our model to identify influential nodes in a user network.

\subsubsection{Baseline comparisons.} To further validate the results of our model, we compared the performance of our model with AMDN-HAGE's \cite{sharmaIdentifyingCoordinatedAccounts2021} performance on this dataset. AMDN-HAGE outputs a matrix of influence scores for each user-user pair, which is similar to the output format of our model. Therefore, all the above experiments can be carried out for AMDN-HAGE. The threshold for AMDN-HAGE was determined by choosing the closest point on the ROC curve to the (0, 1) point for classifying user-user pairs as either semantic-agreed or not. The threshold turned out to be 0.02 for both stance-agreeability and topic-agreeability.

Out of $^{800}{C}_2$ user pairs, AMDN-HAGE identified 122823 pairs of users. Out of the identified pairs, 61943 (50.4\%) pairs were stance-agreed and 55237 (44.9\%) pairs were topic-agreed. More results for different threshold values are demonstrated in Figure \ref{fig:prec-recall-stance-topic}. It can be clearly observed that CCM surpasses AMDN-HAGE in terms of detecting relevant semantically similar user-user pairs by only training on time series data.


\section{Conclusion}
\label{sec:conclusion}
In this work, we proposed an approach to identify causally linked coordinating user pairs by employing convergent cross mapping of their activity traces. We consider a coordinated community as a dynamic system of variables devoid of external influences. The clustered nature of the influence graphs motivated us to pre-cluster users as a preliminary step before applying CCM, thereby reducing the overall search space and addressing RQ2. In conclusion, CCM demonstrates competitive performance in detecting coordinated users on Twitter, particularly excelling in precision answering RQ1. Its ability to identify influential users and causal relationships between users' activities in a community offers a unique advantage over traditional content-based or network-based methods. Further, we showed that our model has the ability to identify influential users in a COVID-19 related dataset.

While our approach offers significant advantages, it also comes with several limitations that need to be addressed (RQ3). First, the computational complexity of analyzing all pairs of users using CCM is substantial, particularly as the number of users increases. This quadratic growth in computational requirements can make it challenging to scale the method for large datasets. Second, the sensitivity to parameter selection, such as embedding dimensions and time delays, requires careful tuning to ensure accurate causal inference. Third, real-time detection of coordinated behaviour is another challenge. The computational demands and the need for timely analysis require efficient algorithms. A study in the evolution and the dynamics of influence graphs could benefit real-time detection. The future direction of this research aims to address the above limitations.




\bibliography{aaai24}

\section{Ethics Checklist}

\begin{enumerate}
    \item For most authors...
    \begin{enumerate}
        \item  Would answering this research question advance science without violating social contracts, such as violating privacy norms, perpetuating unfair profiling, exacerbating the socio-economic divide, or implying disrespect to societies or cultures?
        \answerYes{Yes, this research tries to promote autonomy of users on social media by detecting mass campaigns.}
      \item Do your main claims in the abstract and introduction accurately reflect the paper's contributions and scope?
        \answerYes{Yes, the content of the paper is outlined in the abstract.}
       \item Do you clarify how the proposed methodological approach is appropriate for the claims made? 
        \answerYes{Yes, the results are verified in Sections \ref{subsec:baseline-comparisons}, \ref{subsec:leader-follower}, and \ref{subsec:uncovering}.}
       \item Do you clarify what are possible artifacts in the data used, given population-specific distributions?
        \answerYes{Yes, they are discussed in Sections \ref{subsec:data} and \ref{subsec:baseline-comparisons}.}
      \item Did you describe the limitations of your work?
        \answerYes{Yes, the limitations in runtime, recall, and scalability are discussed at the end of Section \ref{subsec:optimizations} and Section \ref{subsec:baseline-comparisons}.}
      \item Did you discuss any potential negative societal impacts of your work?
        \answerNo{No, because we do not see a potential negative impact of our work on the society.}
          \item Did you discuss any potential misuse of your work?
        \answerNo{No, because we do not see a potential of such misuse of our work.}
        \item Did you describe steps taken to prevent or mitigate potential negative outcomes of the research, such as data and model documentation, data anonymization, responsible release, access control, and the reproducibility of findings?
        \answerYes{}
      \item Have you read the ethics review guidelines and ensured that your paper conforms to them?
        \answerYes{Yes, we did.}
    \end{enumerate}
    
    \item Additionally, if your study involves hypotheses testing...
    \begin{enumerate}
      \item Did you clearly state the assumptions underlying all theoretical results?
        \answerNA{NA}
      \item Have you provided justifications for all theoretical results?
      \answerNA{NA}
      \item Did you discuss competing hypotheses or theories that might challenge or complement your theoretical results?
      \answerNA{NA}
      \item Have you considered alternative mechanisms or explanations that might account for the same outcomes observed in your study?
      \answerNA{NA}
      \item Did you address potential biases or limitations in your theoretical framework?
      \answerNA{NA}
      \item Have you related your theoretical results to the existing literature in social science?
      \answerNA{NA}
      \item Did you discuss the implications of your theoretical results for policy, practice, or further research in the social science domain?
      \answerNA{NA}
    \end{enumerate}
    
    \item Additionally, if you are including theoretical proofs...
    \begin{enumerate}
      \item Did you state the full set of assumptions of all theoretical results? 
      \answerNA{NA}
        \item Did you include complete proofs of all theoretical results?
        \answerNA{NA}
    \end{enumerate}
    
    \item Additionally, if you ran machine learning experiments...
    \begin{enumerate}
      \item Did you include the code, data, and instructions needed to reproduce the main experimental results (either in the supplemental material or as a URL)?
        \answerYes{Yes, the instructions are given in the article.}
      \item Did you specify all the training details (e.g., data splits, hyperparameters, how they were chosen)?
        \answerYes{Yes, the experimental setup is provided in Section \ref{subsec:experimental-setup}.}
         \item Did you report error bars (e.g., with respect to the random seed after running experiments multiple times)?
        \answerNo{No, because the core machine learning component (KNN classifier) is a deterministic algorithm.}
        \item Did you include the total amount of compute and the type of resources used (e.g., type of GPUs, internal cluster, or cloud provider)?
        \answerYes{Yes, that is provided in Section \ref{subsec:results}.}
         \item Do you justify how the proposed evaluation is sufficient and appropriate to the claims made? 
        \answerNo{No, the evaluation method is a core component in the CCM methodology. We do not see the importance of repeating the justification to it in this article.}
         \item Do you discuss what is ``the cost`` of misclassification and fault (in)tolerance?
        \answerYes{Yes, we discuss the improvement in precision. Our work significantly increases precision in identifying coordinated users compared to baselines. Therefore, reducing the cost of misclassifications is a key contribution in our work.}
      
    \end{enumerate}
    
    \item Additionally, if you are using existing assets (e.g., code, data, models) or curating/releasing new assets, \textbf{without compromising anonymity}...
    \begin{enumerate}
      \item If your work uses existing assets, did you cite the creators?
        \answerNo{No, because we use common Python libraries such as Numpy, Scikit-Learn, and Pandas.}
      \item Did you mention the license of the assets?
        \answerNA{NA}
      \item Did you include any new assets in the supplemental material or as a URL?
      \answerNA{NA}
      \item Did you discuss whether and how consent was obtained from people whose data you're using/curating?
        \answerNo{No, because we comply with Twitter Terms of Service.}
      \item Did you discuss whether the data you are using/curating contains personally identifiable information or offensive content?
        \answerNo{No, we do not expose details that can uniquely identify a user on Twitter.}
    \item If you are curating or releasing new datasets, did you discuss how you intend to make your datasets FAIR (see \citet{fair})?
    \answerNA{Not applicable since we do not release our dataset. That would be a violation of Twitter's Terms of Service.}
    \item If you are curating or releasing new datasets, did you create a Datasheet for the Dataset (see \citet{gebru2021datasheets})? 
    \answerNA{NA}
    \end{enumerate}
    
    \item Additionally, if you used crowdsourcing or conducted research with human subjects, \textbf{without compromising anonymity}...
    \begin{enumerate}
      \item Did you include the full text of instructions given to participants and screenshots?
      \answerNA{NA}
      \item Did you describe any potential participant risks, with mentions of Institutional Review Board (IRB) approvals?
      \answerNA{NA}
      \item Did you include the estimated hourly wage paid to participants and the total amount spent on participant compensation?
      \answerNA{NA}
       \item Did you discuss how data is stored, shared, and deidentified?
       \answerNA{NA}
    \end{enumerate}
    
    \end{enumerate}

\end{document}